\documentclass[12pt]{article}
\usepackage[cp1251]{inputenc}
\usepackage{amsmath,amsfonts,euscript,amssymb}
\DeclareSymbolFont{bfitletters}{OML}{cmm}{bx}{it}
\DeclareSymbolFont{bfitoperators}   {OT1}{cmr} {m}{n}
\DeclareMathSymbol{\bfitomega}{\mathord}{bfitletters}{"21}
\DeclareMathSymbol{\bfitrho}{\mathord}{bfitletters}{"1A}
\DeclareMathSymbol{\bfitLambda}{\mathord}{operators}{"03}
\DeclareMathSymbol{\lambda}{\mathord}{letters}{"15}
 \usepackage{epsfig}
 \newcommand{\be}{\begin{equation}}
 \newcommand{\ee}{\end{equation}}
 \newcommand{\bea}{\begin{eqnarray}}
 \newcommand{\eea}{\end{eqnarray}}

\begin{document}

\vskip 0.5cm
\begin{center}
{\bf HIDDEN SYMMETRIES IN MIXMASTER-TYPE UNIVERSE}\\
\vskip 0.5cm
{\bf Alexander Pavlov}
\vskip 0.2cm
{\it Bogoliubov~Laboratory~for~Theoretical~Physics,~Joint~Institute~of~Nuclear~Research,
~Joliot-Curie~str.~6,~Dubna,~141980,~Russia;} \\
{\it Institute of Mechanics and Energetics,\\
Russian State Agrarian University -- Moscow Timiryazev Agricultural Academy,
Timiryazevskaya str. 49, Moscow, 127550 Russia}\\
alexpavlov60@mail.ru
\end{center}
\vskip 0.5cm
A model of multidimensional mixmaster-type vacuum universe is considered. It belongs to a class of pseudo-Euclidean chains characterized by root vectors. An algebraic approach of our investigation is founded on construction of Cartan matrix of the spacelike root vectors in Wheeler -- DeWitt space. Kac -- Moody algebras can be classified according to their Cartan matrix. By this way a hidden symmetry of the model considered is revealed. It is known, that gravitational models which demonstrate chaotic behavior are associated with hyperbolic Kac -- Moody algebras. The algebra considered in our paper is not hyperbolic. The square of  Weyl vector is negative. The mixmaster-type universe is associated with a simply-laced Lorentzian Kac -- Moody algebra. Since the volume of the configuration space is infinite, the model is not chaotic.
\vskip 0.5cm

\vskip 1cm
\begin{flushright}
{\it In honor of the 97th birthday of Louis W. Witten}
\end{flushright}

\section{Introduction}
A chaotic behavior of the mixmaster model was demonstrated in a billiard representation of the dynamics by Misner and Chitre \cite{MisnerDet}.
Chaos is related to the motion in the 2-dimensional hyperbolic space with finite billiard table volume.
Further, it was shown that a wide class of multidimensional cosmological models had a property of stochasticity \cite{IM}. Mixmaster-type models belong to pseudo-Euclidean Toda chains. Bogoyavlenskii introduced generalized Toda chains and discovered that every simple Lie algebra corresponds to a completely integrable generalized Euclidean Toda chain \cite{BogCom}. The equations of motion of particles of these chains admit Lax presentation with an involutive system of integrals.
An ordinary Toda chain corresponds to a simple Lie algebra $\mathfrak{sl}(n, \mathbb{R})$. Hence, Bogoyavlenskii discovered hidden symmetries of the integrable chains and pointed out that a Coxeter group is infinite for cosmological model Bianchi $\rm{IX}$.

A generalized Adler -- van Moerbeke formula for getting Kovalevskaya exponents characterizing generalized Toda chains was obtained in \cite{PavlovReg}.
The necessity of application a classification scheme of noncompact Lie algebras by analogy with using Cartan classification of semi-simple Lie algebras was pointed out \cite{PavlovReg}. Further, it was making clear that hidden symmetries of various gravitational theories are described by infinite dimensional Lie algebras \cite{Sophie}. Gravitational models with chaotic behavior are associated with hyperbolic Kac -- Moody algebras.
A modern review of intimate connection between some gravitational models and hyperbolic Kac -- Moody algebras can be found in \cite{Henneaux,Bel}. Models provided realizations of canonical Lorentzian extensions of all the finite dimensional Lie algebras were demonstrated in \cite{DBuyl}. Their asymptotic dynamics in the vicinity of a spacelike singularity are billiard motion in the fundamental Weyl chamber of the corresponding Kac -- Moody algebras.

Gravitational theories with oscillator behavior regime near a singularity in dimensions $d<10$ are associated with hyperbolic Kac -- Moody algebras, id est hyperbolicity implies chaos \cite{Henneaux}.
There are multidimensional vacuum cosmological models as product of manifolds
$R^1\times M_1\times\ldots\times M_n$ describing Kasner-like behavior near a singularity.
Full classification of homogeneous multidimensional Kaluza -- Klein models is presented in \cite{Dem, Sz}.
Models have been investigated by numerical methods \cite{Witten} and explicit integration \cite{Gav}. Our goal consists in analyzing the model \cite{Witten} that deserved special attention by modern algebraic approach.

\section{Mixmaster model with geometry $R^1\times S^3\times S^3\times S^3$}

A model of a spatially homogeneous vacuum universe with geometry $R^1\times S^3\times S^3\times S^3$ was
introduced and studied in \cite{Witten}. The spacetime metric is defined by an interval
\be
ds^2=-N^2dt^2+e^{2\alpha}\sum\limits_{i=1}^9e^{2\beta_{ij}}\sigma^i\sigma^j,
\ee
where differential forms satisfy the relations
$$
d\sigma^1=\sigma^2\wedge\sigma^3,\qquad d\sigma^4=\sigma^5\wedge\sigma^6,\qquad d\sigma^7=\sigma^8\wedge\sigma^9
$$
with cyclic permutations. Generators of spatial symmetry of a model form a Lie algebra which is a sum of semi-simple algebras
$so(3)\oplus so(3)\oplus so(3)$.
A variety of three-dimensional spheres are orbits of the group of symmetry. Here, $N, \alpha$, and $\beta_{ij}$ are functions of coordinate time $t$ only.
The diagonal matrix $(\beta)_{ij}$ is constructed with entries
\bea
\beta_{11}&=&\frac{2\theta}{\sqrt{3}}+\beta_{+}+\sqrt{3}\beta_{-},\nonumber\\
\beta_{22}&=&\frac{2\theta}{\sqrt{3}}+\beta_{+}-\sqrt{3}\beta_{-},\nonumber\\
\beta_{33}&=&\frac{2\theta}{\sqrt{3}}-2\beta_{+};\nonumber
\eea
\bea
\beta_{44}&=&-\frac{\theta}{\sqrt{3}}-\eta+\psi_{+}+\sqrt{3}\psi_{-},\nonumber\\
\beta_{55}&=&-\frac{\theta}{\sqrt{3}}-\eta+\psi_{+}-\sqrt{3}\psi_{-},\nonumber\\
\beta_{66}&=&-\frac{\theta}{\sqrt{3}}-\eta-2\psi_{+};\nonumber
\eea
\bea
\beta_{77}&=&-\frac{\theta}{\sqrt{3}}+\eta+\varphi_{+}+\sqrt{3}\varphi_{-},\nonumber\\
\beta_{88}&=&-\frac{\theta}{\sqrt{3}}+\eta+\varphi_{+}-\sqrt{3}\varphi_{-},\nonumber\\
\beta_{99}&=&-\frac{\theta}{\sqrt{3}}+\eta-2\varphi_{+}.\nonumber
\eea

\begin{figure}[tbp]
\begin{center}
\includegraphics[width=2.5in]{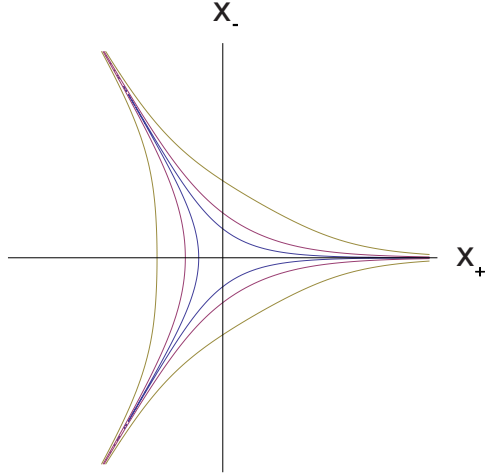}
\caption{\small Lines of level of the function $g(x_{+},x_{-})$ restrict noncompact regions of finite volume with
triangular symmetry and three corners. The angles of the triangles are equal to zero, their sides are parallel.}
\label{Contoursx}
\end{center}
\end{figure}

The super-Hamiltonian defined in the foliated spacetime in Misner's coordinates
${\bf q}(\alpha, \beta_{\pm}, \theta, \psi_{\pm}, \eta, \varphi_{\pm})$ is the following
\be\label{H}
{\cal H}=-\frac{p_\alpha^2}{24}+\sum\limits_{j=1}^8\frac{p_j^2}{2}+\frac{e^{16\alpha}}{2}
\left[
e^{-4\theta /\sqrt{3}}g(\beta)+e^{2\theta /\sqrt{3}}
\left(e^{2\eta}g(\psi)+e^{-2\eta}g(\varphi)\right)
\right],
\ee
where $g(x)$ means a function
\be\label{g}
g(x)\equiv e^{4x_{+}+4\sqrt{3}x_{-}}+e^{4x_{+}-4\sqrt{3}x_{-}}+e^{-8x_{+}}-2e^{4x_{+}}
-2e^{2x_{+}+2\sqrt{3}x_{-}}-2e^{2x_{+}-2\sqrt{3}x_{-}}.
\ee
The potential function of the super-Hamiltonian (\ref{H}) is presented as a sum of three terms
\be\label{U}
U=\frac{1}{2}e^{16\alpha-4\theta/\sqrt{3}}g(\beta)+\frac{1}{2}e^{16\alpha+2\theta/\sqrt{3}+2\eta}g(\psi)+
\frac{1}{2}e^{16\alpha+2\theta/\sqrt{3}-2\eta}g(\varphi).
\ee
Lines of level of the function $g(x_{+},x_{-})$ are depicted in Fig.\ref{Contoursx}.
An asymptotic form is defined by a leading term in (\ref{g})
\bea
&&g(x)\sim e^{-8x_{+}},\qquad x_{+}\to -\infty,\nonumber\\
&&g(x)\sim x_{-}^2e^{4x_{+}},\quad\quad x_{+}\to +\infty,\nonumber
\eea
which defines a potential wall. In force of triangular symmetry of the function $g(x)$ one obtains the next two exponential walls.
Consider a potential wall, when $\beta_{+}=\psi_{+}=\varphi_{+}\equiv x_{+}$ under $x_{+}\to+\infty$. Then, the potential (\ref{U}) takes the form
$$
U=\frac{1}{2}e^{16\alpha}\left[e^{-4\theta/\sqrt{3}}\beta_{-}^2+e^{2\theta/\sqrt{3}}\left(e^{2\eta}\psi_{-}^2+e^{-2\eta}\varphi_{-}^2\right)
\right]x_{-}^2 e^{4x_{+}}.
$$
If, additionally, $\beta_{-}=\psi_{-}=\varphi_{-}\equiv x_{-}$, then
\be
U=\frac{1}{2}e^{16\alpha}\left[e^{-4\theta/\sqrt{3}}+2e^{2\theta/\sqrt{3}}{\rm cosh} 2\eta\right]x_{-}^2 e^{4x_{+}}.
\ee
we see that the pocket under $x_{-}\to 0$ is a three-dimensional space $(\theta, \eta, x_{-})$. So, the volume of the configuration space in indefinite.

The problem is able to be considered as a generalized pseudo-Euclidean Toda chain of exponentially interacted points. Its potential energy is presented as a sum of 18 exponential terms
$$
U ({\bf q})=\sum\limits_{j=1}^{18}c_j\exp {({\bf a}_j, {\bf q})}.
$$
Root vectors in 9-dimensional Wheeler -- DeWitt space
${\bf q}(\alpha, \beta_{\pm}, \theta, \psi_{\pm},\eta,\varphi_{\pm})$
with WDW metric of Lorentzian signature
\be\label{WDWmetric}
(G)_{\mu\nu}={\rm diag} \left(-1/{12},1\ldots,1\right)
\ee
are the following:
\bea
&&{\bf a}_1 (16,4,4\sqrt{3},-4/\sqrt{3},0,0,0,0,0),\nonumber\\
&&{\bf a}_2 (16,4,-4\sqrt{3},-4/\sqrt{3},0,0,0,0,0),\nonumber\\
&&{\bf a}_3 (16,-8,0,-4/\sqrt{3},0,0,0,0,0),\nonumber\\
&&{\bf a}_4 (16,4,0,-4/\sqrt{3},0,0,0,0,0),\nonumber\\
&&{\bf a}_5 (16,2,2\sqrt{3},-4/\sqrt{3},0,0,0,0,0),\nonumber\\
&&{\bf a}_6 (16,2,-2\sqrt{3},-4/\sqrt{3},0,0,0,0,0);\nonumber
\eea
\bea
&&{\bf a}_7 (16,0,0,2/\sqrt{3},4,4\sqrt{3},-2,0,0),\nonumber\\
&&{\bf a}_8 (16,0,0,2/\sqrt{3},4,-4\sqrt{3},-2,0,0),\nonumber\\
&&{\bf a}_9 (16,0,0,2/\sqrt{3},-8,0,-2,0,0),\nonumber\\
&&{\bf a}_{10} (16,0,0,2/\sqrt{3},4,0,-2,0,0),\nonumber\\
&&{\bf a}_{11} (16,0,0,2/\sqrt{3},2,2\sqrt{3},-2,0,0),\nonumber\\
&&{\bf a}_{12} (16,0,0,2/\sqrt{3},2,-2\sqrt{3},-2,0,0),\nonumber
\eea
\bea
&&{\bf a}_{13} (16,0,0,2/\sqrt{3},0,0,2,4,4\sqrt{3}),\nonumber\\
&&{\bf a}_{14} (16,0,0,2/\sqrt{3},0,0,2,4,-4\sqrt{3}),\nonumber\\
&&{\bf a}_{15} (16,0,0,2/\sqrt{3},0,0,2,-8,0),\nonumber\\
&&{\bf a}_{16} (16,0,0,2/\sqrt{3},0,0,2,4,0),\nonumber\\
&&{\bf a}_{17} (16,0,0,2/\sqrt{3},0,0,2,2,2\sqrt{3}),\nonumber\\
&&{\bf a}_{18} (16,0,0,2/\sqrt{3},0,0,2,2,-2\sqrt{3}).\nonumber
\eea
The next 9 vectors of the set
\be\label{spacelike}
({\bf a}_1, {\bf a}_2, {\bf a}_3; {\bf a}_7, {\bf a}_8, {\bf a}_9; {\bf a}_{13}, {\bf a}_{14}, {\bf a}_{15})
\ee
are spacelike, other isotropic. Isotropic vectors are directed into pockets of the potential $U ({\bf q})$ (see Fig.\ref{Contoursx}).
Compactification schemes from initial Kasner states to a final pocket states, figurative speaking ``clogging'', were investigated by numerical analysis in \cite{Witten}. Further, to make clear qualitative peculiarities of the model behavior we concentrate on the spacelike vectors (\ref{spacelike}).
The isotropic vectors are expressed linearly through them
$${\bf a}_4=\frac{1}{2}({\bf a}_1+{\bf a}_2),\qquad {\bf a}_5=\frac{1}{6}(4{\bf a}_1+{\bf a}_2+{\bf a}_3),\quad
{\bf a}_6=\frac{1}{6}({\bf a}_1+4{\bf a}_2+{\bf a}_3);$$
$${\bf a}_{10}=\frac{1}{2}({\bf a}_7+{\bf a}_8),\qquad {\bf a}_{11}=\frac{1}{6}(4{\bf a}_7+{\bf a}_8+{\bf a}_9),\quad
{\bf a}_{12}=\frac{1}{6}({\bf a}_7+4{\bf a}_8+{\bf a}_9);$$
$${\bf a}_{16}=\frac{1}{2}({\bf a}_{13}+{\bf a}_{14}),\qquad {\bf a}_{17}=\frac{1}{6}(4{\bf a}_{13}+{\bf a}_{14}+{\bf a}_{15}),\quad
{\bf a}_{18}=\frac{1}{6}({\bf a}_{13}+4{\bf a}_{14}+{\bf a}_{15}).$$
The isotropic vectors do not belong to the root lattice of spacelike vectors.

Let us construct a 9-dimensional block-type Gram matrix of the spacelike vectors ${\bf a}_i$ (\ref{spacelike}) with WDW metric (\ref{WDWmetric})
\be\label{Gram}
<{\bf a}_i, {\bf a}_j>\equiv G_{\mu\nu}a_i^\mu a_j^\nu=
24
\left(
\begin{array}{ccc}
A_1&-I&-I\\
-I&A_1&-I\\
-I&-I&A_1
\end{array}
\right).
\ee
The Gram matrix (\ref{Gram}) is built of blocks $\hat{A}_1$ and a degenerated matrix $\hat{I}$
\be\label{Cartanmatrix}
\hat{A}_1:=
\left(
\begin{array}{ccc}
2&-2&-2\\
-2&2&-2\\
-2&-2&2
\end{array}
\right),
\qquad
\hat{I}:=
\left(
\begin{array}{ccc}
1&1&1\\
1&1&1\\
1&1&1
\end{array}
\right).
\ee
The matrix $\hat{A}_1$ corresponds to Cartan matrix of the mixmaster cosmological model \cite{Bel,PavlovG&C}. The Cartan matrix $\hat{A}$ is convenient to be presented as a product of a block diagonal matrix and a symmetric one
\be\label{produ}
\hat{A}:=
\frac{2<{\bf a}_i, {\bf a}_j>}{<{\bf a}_j, {\bf a}_j>}=
\left(
\begin{array}{ccc}
A_1&-I&-I\\
-I&A_1&-I\\
-I&-I&A_1
\end{array}
\right)=
\left(
\begin{array}{ccc}
A_1&0&0\\
0&A_1&0\\
0&0&A_1
\end{array}
\right)
\left(
\begin{array}{ccc}
E&I/2&I/2\\
I/2&E&I/2\\
I/2&I/2&E
\end{array}
\right),
\ee
where $\hat{E}$ is a unit three-dimensional matrix.
The determinant of the matrix $\hat{A}$ is negative:
$${\rm det}(\hat{A})=\left({\rm det}\hat{A}_1\right)^3=(-32)^3.$$
It is a product of determinant of the block diagonal matrix and a unit.
The eigenvalues of the matrix $\hat{A}$ are the following:
$(-8, 4, 4, 4, 4, 4, 4, 1, 1).$ Thus, in force of existence of one negative eigenvalue,
we have proved that the matrix $\hat{A}$ is Lorentzian. At the same time, it is not hyperbolic because of nonzero minors of its diagonal elements: $M_{ii}=-12288$. Minors of diagonal elements of the $s$-th order $(s=2,\ldots, 7)$ are zero or negative:
$M_{ii}^{(2)}=(0, 3)$, $M_{ii}^{(3)}=(-32, -8, 0),$ $M_{ii}^{(4)}=(-112, -64, -48),$
$M_{ii}^{(5)}=(-384, -256),$ $M_{ii}^{(6)}=(-1280, -1024),$ $M_{ii}^{(7)}=(-4096).$

Cartan matrix associated with some root system can define a new complex Lie algebra.
The structure of an algebra is encoded in its Cartan matrix.
By definition \cite{Kac}, a non-degenerated symmetric Cartan matrix is $r\times r$ matrix such that $A_{ii}=2$,
$A_{ij}\le 0$ for $i\ne j$.
Given the Cartan matrix $\hat{A}$, an algebra of Kac -- Moody $\mathfrak{g} (\hat{A})$ is defined as a free Lie algebra generated by the generators $h_i, e_i, f_i$ $(i=1,\ldots,r)$ modulo Serre relations. The generators
satisfy the Chevalley commutation relations
\begin{eqnarray}
&&[h_i, h_j]=0,\qquad\quad [e_i,f_j]=\delta_{ij}h_j,\label{LieI}\\
&&[h_i,e_j]=A_{ij}e_j,~\quad [h_i,f_j]=-A_{ij}f_j,\label{LieII}
\end{eqnarray}
where $\delta_{ij}$ is the Kronecker symbol.
A triangular decomposition of ${\mathfrak g}(\hat{A})$ has a form of direct sum of vector spaces
$$
{\mathfrak g}(\hat{A})={\mathfrak n}_{-}\oplus{\mathfrak h}\oplus{\mathfrak n}_{+}.
$$
Here ${\mathfrak h}$ is Cartan subalgebra, which is formed as abelian subalgebra, spanned by the elements $h_i$.
Its dimension $r$ is the rank of the algebra.
The subspaces ${\mathfrak n}_{-}$, ${\mathfrak n}_{+}$
are freely generated.

In our case considered, the Cartan matrix (\ref{produ}) is symmetrical by its origin, so we deal with a simply-laced algebra. The algebra ${\mathfrak g}(\hat{A})$ is constructed by using 27 generators $h_i, e_i, f_i$ $(i=1,\ldots,9)$ satisfied the Chevalley relations (\ref{LieI}), (\ref{LieII}).
A class of Kac -- Moody algebras, corresponding to the case when the matrix $\hat{A}$
has one negative and other positive eigenvalues is called Lorentzian \cite{Henneaux}.
There is a subclass of the Lorentzian algebras, known as hyperbolic ones.
Kac -- Moody algebra is called hyperbolic if, in addition to Lorentzian,
its Dynkin diagram is that, if removing from it one node one obtains the Dynkin diagram of affine
or finite Kac -- Moody algebra. Affine algebras proved useful in describing various interesting physical problems \cite{Bombay}.

We have proved that the mixmaster-type universe is associated with a Lorentzian Kac -- Moody algebra.
The elements of the algebra, besides of the generators $h_i, e_i, f_i (i=1,\ldots,9)$, are derived by commutators $[e_1,e_2], [e_1,[e_1,e_2]],\ldots$. With accounts of the Serre relations, e.g. $[e_1,[e_1,[e_1,e_2]]]=0$,
$[e_1,[e_1,e_4]]=0$,
the dimension of the algebra is finite.

\begin{figure}[tbp]
\begin{center}
\includegraphics[width=1.5in]{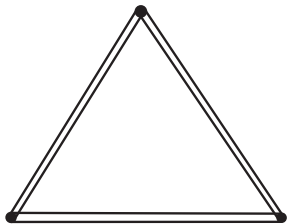}
\caption{\small Dynkin diagram, corresponding to the Kac -- Moody hyperbolic algebra with the Cartan matrix
$\hat{A}_1$ of the mixmaster model (\ref{Cartanmatrix}). There are three nodes and connected them double lines.
The Dynkin diagram corresponding to 9-dimensional Cartan matrix $\hat{A}$ can be obtained with utilizing the three mixmaster diagrams  corresponding to 3-dimensional Cartan matrices $\hat{A}_1$ by drawing all lines connected the all nodes.}
\label{Dynkin7}
\end{center}
\end{figure}

Kovalevskaya exponents of the model considered were calculated in \cite{PavlovG&C}. In addition to integer values, fractional exponents were found, which indicates the existence of chaotic regimes of the model.
A trivial degenerated case
$\theta=\eta=0$;
$\beta_{+}=\psi_{+}=\varphi_{+};$
$\beta_{-}=\psi_{-}=\varphi_{-}$
is reduced to the mixmaster case.
Then the Gram matrix (\ref{Cartanmatrix}) is equivalent to
corresponding Cartan matrix of hyperbolic Kac -- Moody algebra associated to the mixmaster model.
A billiard table is identical to Weyl chamber of the Kac --Moody algebra which Dynkin diagram is drawn in Fig.\ref{Dynkin7}.
Lorentzian Kac -- Moody algebras are unclassified yet, while all hyperbolic algebras are classified \cite{Lisa}, there are not with rank of high than 10. The Dynkin diagram corresponding to 9-dimensional Cartan matrix $\hat{A}$ (\ref{produ}) can be obtained with utilizing the three mixmaster diagrams (see Fig.\ref{Dynkin7}) corresponding to 3-dimensional Cartan matrices $\hat{A}_1$ (\ref{Cartanmatrix})
by drawing all lines connected the all nodes.

Let us outline some correspondences between a billiard system and Weyl group of reflections $W(\hat{A})$ of some Lie algebra \cite{Henneaux}.

$\bullet$
Reflections in hyperplanes orthogonal to simple roots leave the root system invariant. The corresponding finite-dimensional group is called the Weyl group. The Weyl group transforms one Weyl chamber to another.

$\bullet$
Billiards describing the evolution of the universe in the vicinity of the singularity correspond to the Kac -- Moody algebras. Kasner's indices describing the free motion of the universe between reflections from the walls correspond to the elements of the Cartan subalgebra.

$\bullet$
The dominant walls responsible for the transition from one Kasner era to another correspond to the simple roots of the Kac -- Moody algebra.

$\bullet$
The group of reflections in cosmological billiards is the Weyl group of the algebra. The billiard table is identified with the Weyl camera.

$\bullet$
If the billiard region of a gravitational system can be identified with the fundamental Weyl chamber of some hyperbolic Kac -- Moody algebra, then the dynamics of the system is chaotic.

Weyl reflections act on the simple roots by a rule
$$
\hat{w}_i({\bf a}_j)={\bf a}_j-\frac{2<{\bf a}_i, {\bf a}_j>}{<{\bf a}_i, {\bf a}_i>}{\bf a}_i.
$$
The fundamental Weyl chamber $\mathfrak{C}_W$ is a wedge about the origin of root space delimited by the reflecting hyperplanes. A billiard table is identified with Weyl chamber of the algebra.
The fundamental weights $\{\bfitLambda_i\}$ are vectors in the dual space ${\mathfrak h}^*$ of the Cartan subalgebra defined by the scalar products \cite{Bel}
$$
<\Lambda_i, {\bf a}_j>=\frac{1}{2}<{\bf a}_i, {\bf a}_j>\delta_{ij}=24\delta_{ij}.
$$
Thus, one yields
\bea
&&\Lambda_1 (-2,1,\sqrt{3},-{4}/{\sqrt{3}},0,0,0,0,0),\nonumber\\
&&\Lambda_2 (-2,1,-\sqrt{3},-{4}/{\sqrt{3}},0,0,0,0,0),\nonumber\\
&&\Lambda_3 (-2,-2,0,-{4}/{\sqrt{3}},0,0,0,0,0),\nonumber\\
&&\Lambda_4 (-2,0,0,{2}/{\sqrt{3}},1,\sqrt{3},-2,0,0),\nonumber\\
&&\Lambda_5 (-2,0,0,{2}/{\sqrt{3}},1,-\sqrt{3},-2,0,0),\nonumber\\
&&\Lambda_6 (-2,0,0,{2}/{\sqrt{3}},-2,0,-2,0,0),\nonumber\\
&&\Lambda_7 (-2,0,0,{2}/{\sqrt{3}},0,0,2,1,\sqrt{3}),\nonumber\\
&&\Lambda_8 (-2,0,0,{2}/{\sqrt{3}},0,0,2,1,-\sqrt{3}),\nonumber\\
&&\Lambda_9 (-2,0,0,{2}/{\sqrt{3}},0,0,2,-2,0).\nonumber
\eea

Weyl vector ${\bfitrho}$ is defined according to its scalar products \cite{Bel}
with every spacelike root vector ${\bf a}_i$ (\ref{spacelike}):
\be\label{vectorWeyl}
<{\bfitrho}, {\bf a}_i>=\frac{1}{2}<{\bf a}_i, {\bf a}_i>=24.
\ee
The result of calculation presented in component form gives
$$
<{\bfitrho}, {\bf a}_i>\equiv G_{\mu\nu}{\rho}^\mu a_i^\nu = \frac{1}{2}G_{\mu\nu}a_i^\mu a_i^\nu=24.
$$
The Weyl vector can be obtained by summing the fundamental weights \cite{Bel}
\be
\bfitrho=\sum\limits_{i=1}^9\Lambda_i=(-18,0,\ldots,0).
\ee
It has only a timelike component. Its scalar squared is negative:
$$<{\bfitrho}, {\bfitrho}>\equiv G_{\mu\nu}\rho^\mu\rho^\nu=-27<0.$$
Spacelike roots (\ref{spacelike}) have uniform square length $<{\bf a}, {\bf a}>=24$ and lie on a hyperboloid with axis pointing into the timelike direction. 

\section*{Conclusions}

The mixmaster-type vacuum universe with geometry $R^1\times S^3\times S^3\times S^3$ is analyzed by the modern algebraic methods. It belongs to a class of pseudo-Euclidean chains characterized by root vectors.
Cartan matrix of nine spacelike root vectors in Wheeler -- DeWitt (WDW) space was constructed. Kac -- Moody algebras are classified according their Cartan matrices. By this approach a hidden symmetry of the model considered is revealed. It is associated with a simply-laced Lorentzian Kac -- Moody algebra. The discrete Weyl group of reflections belongs to the continuous Lorentz group.
The Weyl vector is directed into timelike direction. Its scalar square is negative. The model presented an interesting case associated with the Lorentzian Kac -- Moody algebra not belonging to hyperbolic ones. In force of indefinite configuration space, the model is not chaotic.

\section*{Acknowledgments}

Author thanks to Profs. M. Henneaux, V. D. Ivashchuk, and M. Szyd{\l}owski for useful discussions and  for acquaintance with their papers.


\end{document}